\documentclass[prb,aps,twocolumn,floatfix]{revtex4}

\usepackage{mathrsfs}
\usepackage{amssymb}
\usepackage{lscape}
\usepackage{graphicx}
\usepackage{tabularx}
\usepackage{subfigure}
\usepackage{epstopdf}
\usepackage[usenames,dvipsnames]{xcolor}

\newcommand{\beq}{\begin{equation}}
\newcommand{\eeq}{\end{equation}}

\begin{document}


\title{Reactions between cold methyl halide molecules and alkali-metal atoms}

\author{Jesse J. Lutz and Jeremy M. Hutson
\thanks{Corresponding author}
\email[e-mail: ]{j.m.hutson@durham.ac.uk}}
\affiliation{Joint Quantum Centre (JQC) Durham-Newcastle,
          Department of Chemistry,
          Durham University,
          South Road, Durham, DH1 3LE,
          United Kingdom}

\begin{abstract}
We investigate the potential energy surfaces and activation energies for
reactions between methyl halide molecules CH$_{3}X$ ($X$ = F, Cl, Br, I) and
alkali-metal atoms $A$ ($A$ = Li, Na, K, Rb) using high-level {\it ab initio}
calculations. We examine the anisotropy of each intermolecular potential energy
surface (PES) and the mechanism and energetics of the only available exothermic
reaction pathway, ${\rm CH}_{3}X+A\rightarrow{\rm CH}_{3}+AX$. The region of
the transition state is explored using two-dimensional PES cuts and estimates
of the activation energies are inferred. Nearly all combinations of methyl
halide and alkali-metal atom have positive barrier heights, indicating that
reactions at low temperatures will be slow.
\end{abstract}
\date{\today}

\maketitle


\section{Introduction}
\label{sec:1} Recent improvements in experimental techniques for cooling
gas-phase atoms and ions to cold ($T < 1$~K) and ultracold ($T < 1$~mK)
temperatures have ushered in an exciting new era of low-energy structure and
dynamics research. Interest has broadened to encompass neutral diatomic and
polyatomic molecules, where fundamental applications are being pursued,
including controlled ultracold chemistry,\cite{Kpccp2008} quantum information
and computing,\cite{Dprl2002} and high-precision measurements that place limits
on the time-dependence of fundamental
constants.\cite{HSTHprl2002,VKBSRMepjd2004,ZBBCLYcpc2008}

Deceleration and trapping of molecules presents a more formidable challenge
than for atoms, due to additional internal vibrational and rotational energy
structure, enhanced long-range forces resulting from molecular multipole
interactions, and the possibility of collision-induced chemical reactivity. In
the cold regime, molecules exist primarily in their ground electronic and
rovibrational states and long-range forces and resonance phenomena play a
dominant role in the outcome of collisions.

The current approaches for producing ultracold molecules fall into two
categories. First, there are {\it direct} methods for cooling, where molecules
already in their desired chemical form are cooled from higher temperatures.
Helium buffer-gas cooling,\cite{WdGFDn1998} Stark
deceleration,\cite{BMirpc2003} and Zeeman
deceleration\cite{VMAMMpra2007,HSAVMpra2007,NPLNCERnjp2007} are the most widely
used techniques of this kind, but laser cooling of SrF\cite{SBDn2010} has also
been reported. Opto-electrical cooling using the Sisyphus effect has also
recently been demonstrated for electrically trapped CH$_{3}$F
molecules.\cite{ZEGPMSvBMRn2012} Secondly, there are {\it indirect} methods,
where previously cooled atoms are combined by
photoassociation\cite{HSirpc2006,JTLJrmp2006} or tuning across magnetic
Feshbach resonances.\cite{HSirpc2006,KGJrmp2006} While indirect methods have
been applied with much success to produce ultracold alkali-metal dimers, direct
methods are currently more generally applicable for other
molecules.\cite{CO,ND3,OH,YbF,H2CO,SO2,NH} The present lower limit for
temperatures that may be accessed using direct methods is 10 to 100 mK, and as
a result ``second-stage'' cooling techniques are needed to bridge the gap to
gain entrance into the microkelvin regime.

Among the most promising second-stage cooling methods are {\it sympathetic
cooling},\cite{SHprl2004} whereby a thermally hot species is cooled by
immersion within a sample of another previously cooled species, and evaporative
cooling\cite{SHYn2012}, where the hottest molecules are selectively removed
from the sample. Sympathetic cooling requires thermalization to occur before
molecules are lost from the trap. Magnetic and electrostatic traps rely on the
atoms and molecules remaining in specific low-field seeking states resulting
from the Zeeman and Stark splittings that exist in an applied field. Inelastic
or reactive collisions that cause transitions away from these states convert
internal energy into translational energy and result in ejection of both
species from the trap. The major challenge is therefore to minimize inelastic
and reactive collisions. If reactive collisions can be ruled out as
energetically forbidden, then it is the ratio of elastic to inelastic cross
sections that determines the likelihood of success of sympathetic cooling.
Inelastic cross sections are often suppressed at low collision energies and
fields by centrifugal barriers.\cite{Volpi:2002}

Symmetric-top molecules have particular advantages for sympathetic cooling.
They have near-first-order Stark effects, which allow them to be decelerated
and trapped electrostatically and then brought into contact with a magnetically
trapped coolant. The fact that the two species are trapped independently allows
the clouds to be matched in size even when the temperatures are different. In
particular, there has been extensive experimental and theoretical work on the
collisions of NH$_3$ and ND$_3$ with Rb.\cite{ZHpra2008,ZHpra2009,PFZHLprl2011}
\.Zuchowski and Hutson \cite{ZHpra2008} explored the potential energy surfaces
(PESs) for NH$_3$ interacting with alkali-metal and alkaline-earth atoms and
found them all to be deep and strongly anisotropic. Among the potentials for
interactions with easily coolable atoms, Rb-NH$_3$ was the least anisotropic,
so this system was chosen for detailed collision calculations.\cite{ZHpra2009}
However, it was found that, even in the absence of an electric field, molecules
that are initially in the upper ($f$) component of the tunneling doublet
undergo fast inelastic transitions to the lower ($e$) component. Parazzoli {\it
et al}.\ \cite{PFZHLprl2011} subsequently carried out an experiment in which an
electrostatic trap containing cold ND$_3$ was overlapped with a magnetic trap
containing Rb, and observed inelastic collisions even faster than predicted;
they also carried out collision calculations in an electric field, and
demonstrated that the field could cause substantial changes in the
inelasticity.

One reason for the fast inelastic collisions involving ND$_3$ in its upper
tunneling is that the kinetic energy release due to tunneling persists even at
zero electric field. The tunneling splitting in ND$_3$ is 0.0534 cm$^{-1}$,
which corresponds to a kinetic energy release of 77~mK. This is considerably
higher than the centrifugal barriers so precludes the possibility of
centrifugal suppression of inelasticity for low-energy collisions. This led us
to consider whether other symmetric-top molecules without tunneling would be
better candidates for sympathetic cooling. In this work we begin to investigate
the prospect of sympathetic cooling of the methyl halides, CH$_{3}X$ ($X$ = F,
Cl, Br, I) by alkali-metal atoms $A$ ($A$ = Li, Na, K, Rb). The methyl halides
all have substantial dipole moments ($\mu = 1.858$, 1.892, 1.822, 1.620 D,
respectively) so are amenable to electrostatic deceleration and
trapping.\cite{CH3F} However, there is a considerable class of reactions
between alkali-metal atoms and halogen-containing molecules for which the
reaction pathways are barrierless\cite{DFahv1979} or have barriers submerged
beneath the energy of separated reactants. In the context of sympathetic
cooling, it is crucial to rule out the possibility of fast reactions between
the colliding species before considering nonreactive scattering phenomena.

Reactive collisions between CH$_{3}X$ and $A$ at cold and ultracold
temperatures are likely only if the reactions are exothermic and are either
barrierless or have submerged barriers. For the species of interest here, there
is only one exothermic reaction pathway, a dissociative charge transfer (DCT)
forming methyl radical and alkyl halide products: ${\rm CH}_{3}X + A
\rightarrow {\rm CH}_{3} + AX$. The primary goal of the present study is to
determine whether activation barriers exist for this class of reactions.

Reactions between alkali-metal atoms and methyl halides have been studied
intensively in the field of reaction dynamics. However, very few {\it ab
initio} studies have pursued gas-phase activation barriers for these reactions.
Chang {\it et al}.\cite{CEHPWjcp1997} and Hudson {\it et al}.\cite{HNOPRfd2001}
studied ground-state and excited-state potential energy surfaces for Li +
CH$_{3}$F and Na + CH$_{3}X$ ($X$ = F, Cl, Br), respectively, but focused on
the regions around the global minima. They did not characterize transition
states and indeed they did not find the surfaces that are important for the
reactions: evidently the reactive surfaces correspond to high-lying excited
states at near-equilibrium geometries and fall rapidly in energy as the C-$X$
bond is stretched. We could not find in the literature any study which located
transition states for the DCT reactions.

Thorough {\it ab initio} studies do exist modeling the related dissociative
electron attachment (DEA) reactions, ${\rm CH}_{3}X + {\rm e}^{-} \rightarrow
{\rm CH}_{3} + X^{-}$, for
CH$_{3}$F,\cite{Wjpc1979,HREjcs1986,BGMSjacs1992,Pjms1997}
CH$_{3}$Cl,\cite{Wjpc1979,HREjcs1986,BBBRTcpl1989,SSTjcp2002} and the remaining
methyl halides.\cite{Wjpc1979,BGMSjacs1992,AWLMjcp2004} Wu\cite{Wjpc1979}
proposed that CH$_{3}X + A$ DCT reactions could be modeled by their analogous
${\rm CH}_{3} + {\rm e}^{-}$ DEA processes. In his work the alkali-metal atom
was approximated by a free electron and the competing neutral and anionic
potential energy curves along the C-$X$ bond-breaking coordinate were
constructed semi-empirically. These two curves cross at a point whose energy
can be viewed as the activation energy for the DEA reaction. This activation
energy was found to have a strong dependence on the identity of the halogen,
with the predicted values decreasing from 1.90 eV for CH$_{3}$F to 0.026 eV for
CH$_{3}$I. Polarization and steric effects due to the presence of the
alkali-metal are however a major concern in the DCT systems and it is
questionable whether accurate reaction energetics can be obtained within this
approximation. On the other hand, it is useful to understand when such
approximations are valid, since simple models are computationally less taxing
than the conventional supermolecular approach.

The effect of the direction of approach on reactive collisions of alkali-metal
atoms with symmetric-top molecules has been explored experimentally. The groups
of Brooks,\cite{Bjcp1969,MBjacs1975,Bs1976,BMPcpl1979,BHPCjpc1992}
Bernstein,\cite{BBjcp1969,PCBjpc1981,PCBcpl1982,Bjcp1985,PBarpc1989,GBjcp1990}
and Stolte\cite{Sbbpc1982,Sambm1988,JPSjpc1997,BSjpc1997,BMJSjcp1991} oriented
molecules in hexapole electric fields, while the group of Loesch oriented them
in high static electric fields.\cite{LRjpc1991} This work helped classify the
CH$_{3}$I + $A$ DCT reactions as ``rebound'' reactions, initiated by a close,
orientation-dependent approach of the alkali-metal atom, where one reactant
must hit the other more-or-less head-on for reaction to occur. There is a
strong propensity for backward scattering of the alkyl halide product. This
dependence of the probability of electron transfer on the molecular orientation
is characterized by the ``acceptance angle''; see Ref.\
\onlinecite{MSGVirpc2003} for a review.

Wiskerke {\it et al}.\cite{WSLLpccp2000} revisited the experiments on the
CH$_3$I + K $\rightarrow$ CH$_3$ + KI reaction more recently. They argued that
the reaction is more likely to proceed if the collision time is comparable to
or longer than the time required for the C-I bond to stretch. They concluded
their study by calling for further theoretical examination of (1) the extent to
which the CH$_{3}$I symmetric stretch is coupled to the relative motion in the
entrance valley, (2) the topology of the seam between the covalent and ionic
potentials, and (3) the general mechanism whereby CH$_{3}$I comes to act as a
charge receptor.

The present work is motivated both by interest in cold molecular collisions and
by the theoretical questions regarding the mechanism and energetics of the
CH$_{3}X + A$ DCT reactions. We perform calculations that explicitly model the
approach of the alkali-metal atom in order to obtain an approximate activation
barrier for the reaction pathway of each system. This paper is structured as
follows: In Sec.\ \ref{sec:2} we introduce the computational methods applied
throughout the study. In Sec.\ \ref{sec:3a} we consider nonreactive
intermolecular potentials, characterizing important stationary points and
comparing anisotropies. Having established the most favorable intermolecular
orientation for reaction, we then investigate the topology of the ground-state
reactive PES in Sec.\ \ref{sec:3b}. In Sec.\ \ref{sec:3c} we explore
minimum-energy reaction profiles and estimate the activation energy for each
system. Finally, in Sec.\ \ref{sec:4} we summarize the implications of the
results in the context of sympathetic cooling and suggest particularly
promising sympathetic cooling partners upon which to focus in future work.

\section{Computational methods}
\label{sec:2}

The interaction energy of two monomers $A$ and $B$ is defined as $E^{AB}_{\rm
int} = E^{AB} - E^{A} - E^{B}$, where $E^{AB}$ is the energy of the dimer and
$E^{A}$ and $E^{B}$ are the energies of the isolated monomers. For potential
energy surfaces between rigid monomers, we use the single-reference
coupled-cluster (CC) method including single and double excitations and a
noniterative treatment of triple excitations, abbreviated as CCSD(T). In
particular, we use the partially spin-restricted open-shell CCSD(T) method,
RCCSD(T),\cite{KHWjcp1993,WGBjcp1993} because it offers a highly accurate
treatment of dynamical correlation at a relatively low computational cost,
while avoiding potential spin-contamination issues often associated with
unrestricted variants. All correlation energy calculations in this study were
performed with core orbitals kept frozen.

The RCCSD(T) method can produce divergent energetics when nondynamical
correlation effects become important, which can occur for stretched nuclear
configurations. One solution to this is to use multireference approaches such
as the complete-active-space self-consistent field (CASSCF) method, which
always gives qualitatively correct energetics for reactive surfaces, provided
the active space is adequately large.\cite{WKjcp1985,KWcpl1985} In the present
work, we use CASSCF calculations with 11 electrons distributed among 10
orbitals, designated (11,10); these orbitals correspond asymptotically to the
valence $s$ and $p$ shells of the alkali-metal atom and the two higher-lying
A$'$ and the two E valence molecular orbitals of CH$_{3}X$.

The CASSCF approach does not provide accurate relative energetics and can
produce artificial transition states if the active space is inadequate; see
Ref.\ \onlinecite{MAGLPijck2012} for a recent example. Many of the shortcomings
associated with CASSCF may be overcome by applying a multi-state
multi-reference second-order perturbation theory treatment of the correlation
energy (MS-MR-CASPT2) on top of a state-averaged CASSCF
reference.\cite{Wmp1996,FMRcpl1998} When employing the more expensive
MS-MR-CASPT2 approach, we use a smaller and computationally more tractable
(3,6) active space, which differs from the (11,10) active space by the omission
of the two valence CH$_{3}X$ molecular orbitals of E symmetry. A level shift of
0.2 was also applied to avoid intruder state problems.\cite{RAcpl1995}

Another method which has been shown to provide accurate relative energetics for
potential energy surfaces involving cleavage of a single bond is the rigorously
size-extensive completely renormalized CC method with singles, doubles, and
non-iterative triples, referred to as
CR-CC(2,3).\cite{PWjcp2005,PWGKcpl2006,WLPGmp2006,crcc23open}  This method has
been shown to be as accurate as RCCSD(T) is situations where the latter
performs well,\cite{ZGLWPTjcp2008,ZTGLLPTjpca2009} while succeeding in a few
specific cases where RCCSD(T) fails, such as for single-bond
breaking.\cite{crccl_ijqc,crccl_ijqc2} When computing reaction barrier heights,
experience has shown that when CCSD(T) and CR-CC(2,3) agree, both faithfully
reproduce full CCSDT results.\cite{ZGLWPTjcp2008,ZTGLLPTjpca2009} Thus, in this
work we use the CR-CC(2,3) method as a diagnostic tool for testing the accuracy
of CCSD(T).

First-, second- and third-row atoms are described using Dunning's cc-pV$x$Z or
aug-cc-pV$x$Z basis sets\cite{Djcp1989,KDHjcp1992} (where $x$ is the cardinal
number of the basis set), abbreviated throughout as V$x$Z or AV$x$Z,
respectively. Where Dunning's basis sets are too large to be computationally
tractable, Pople's 6-31G* basis
sets\cite{HDPjcp1972,DPjcp1975,HPtca1973,FPHjcp1982} are used instead. We use
the Stuttgart ECP10MDF pseudo-potentials (PPs) for K and Br and the ECP28MDF
PPs for Rb and I.\cite{LSMSjcp2005,PFGSDjcp2003,PSFSjpca2006} We use the usual
complementary basis sets for K and Rb\cite{LSMSjcp2005} and the
ECP10MDF\_AV$x$Z\cite{PFGSDjcp2003} and ECP28MDF\_AV$x$Z PPs basis sets for Br
and I, respectively; the cardinal number $x$ is chosen to match that for the
all-electron V$x$Z or AV$x$Z basis sets used for the other atoms in the same
calculation.

Some regions of the potential energy surfaces computed here are dominated by
van der Waals forces. The representation of the dispersion energy is greatly
improved by inclusion of midbond functions and elimination of basis set
superposition error. For nonreactive surfaces, we include midbond functions
with exponents {\it sp}: 0.9, 0.3, 0.1 for the AVDZ basis set and,
additionally, {\it df}: 0.6, 0.2 for the AVTZ basis set and correct for
basis-set superposition error using the counterpoise correction.\cite{BBmp1970}
For reactive surfaces, we use results from calculations without counterpoise
corrections because they can sometimes worsen results in such
situations.\cite{SSjcc1997}

In this work, single-point energy calculations are often preceded by geometry
optimizations, allowing secondary geometrical parameters to relax in response
to those explicitly varied. As an example, an optimization using restricted
second-order M{\o}ller-Plesset perturbation theory (RMP2) with the AVDZ basis
set and followed by an RCCSD(T) single-point energy calculation using the AVTZ
basis set is designated by the abbreviation MP2/AVDZ//RCCSD(T)/AVTZ. If the two
basis sets are identical, only one is given.

The geometrical parameters that are always allowed to vary during optimizations
are the C-H internuclear distance $R_{\rm CH}$, the $X$-C-H bond angle
$\theta_{ X{\rm CH}}$, and, where applicable, the $A$-$X$ internuclear distance
$R_{AX}$ and the $X$-C-$A$ bond angle, $\theta_{\rm XCA}$. Unless otherwise
noted, the CH$_{3}X$ fragment is always restricted to $C_{\rm 3v}$ symmetry and
the $A$ fragment is constrained to approach from a H-C-$X$-$A$ torsion angle of
$\phi=180^{\circ}$.

The RMP2, RCCSD(T), CASSCF, and MS-MR-CASPT2 calculations were performed with
MOLPRO\cite{Molpro} and CR-CC(2,3) calculations were performed using
GAMESS.\cite{GAMESS1,GAMESS2} Basis sets and PPs were retrieved from the EMSL
basis set exchange\cite{EMSL} and Stuttgart/Cologne Group PP
repository,\cite{Stuttgart} respectively.

\section{Results and Discussion}

\begin{figure*}
\includegraphics[width=6.69in]{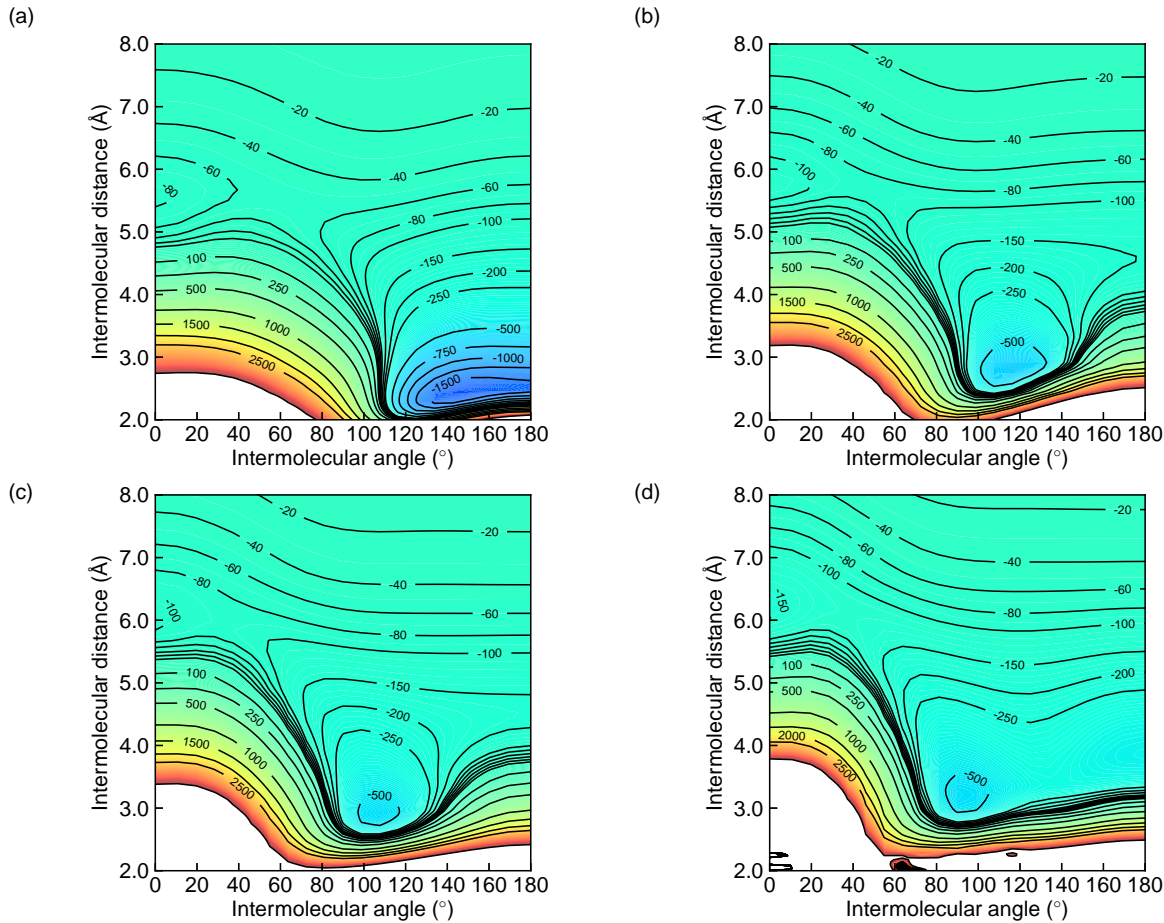}
\caption{ \label{fig:int} Nonreactive RCCSD(T)/AVDZ potential energy surfaces
between Li and (a) CH$_{3}$F, (b) CH$_{3}$Cl, (c) CH$_{3}$Br, and (d) CH$_{3}$I.
Contours represent interaction energies in cm$^{-1}$.}
\end{figure*}

\subsection{Nonreactive intermolecular potential energy surfaces for CH$_{3}X + A$}
\label{sec:3a}

Potential energy surfaces representing the nonreactive interaction between Li
and CH$_{3}$F, CH$_{3}$Cl, CH$_{3}$Br, and CH$_{3}$I are shown in Figures
\ref{fig:int}a to \ref{fig:int}d, respectively. The surfaces are constructed at
the RCCSD(T)/AVDZ level of theory on a grid of points composed of polar angles
$\theta$ corresponding to a 21-point Gauss-Lobatto quadrature and a regularly
spaced set of 21 intermolecular separations $R$, forming a grid comprised of
441 points. The polar angles $\theta=0^{\circ}$ and $\theta=180^{\circ}$
correspond to approach of the Li to the side of the molecule closest to the
H-H-H plane and to the halogen atom, respectively. The monomer geometries are
held fixed at their experimentally determined equilibrium geometries.\cite{CRC}

Two minima are evident on the CH$_{3}$F + Li surface in Figure \ref{fig:int}a.
The first minimum occurs when Li is positioned at $\theta=0^{\circ}$, $6$\ \AA\
away from the molecular center of mass. The complex has $C_{\rm 3v}$ symmetry
in this configuration. Given the rather large intermolecular separation and
relatively weak attraction ($\sim 100\ {\rm cm}^{-1}$), this local minimum may
be attributed primarily to induction and dispersion forces. The other minimum
is located at $\theta=140^{\circ}$, 2.5 \AA\ away from the molecular center of
mass. The symmetry of the complex in this orientation is $C_{\rm s}$. By
symmetry there are actually three minima of this type, each bound by $>2000\
{\rm cm}^{-1}$, making them the global minima. The CH$_{3}$F + Li surface
exhibits strong anisotropy on the repulsive wall, with the inner turning point
rapidly receding by $\sim2.5$ \AA\ between intermolecular angles of
$\theta=50^{\circ}$ and $\theta=100^{\circ}$. In the region with
$\theta>120^{\circ}$, the strong covalent attraction responsible for the global
minima becomes virtually independent of angle, with a relatively low saddle
point occurring at $\theta=180^{\circ}$.

The CH$_{3}$Cl + Li, CH$_{3}$Br + Li, and CH$_{3}$I + Li surfaces are shown in
Figures \ref{fig:int}b, \ref{fig:int}c, and \ref{fig:int}d, respectively. The
two minima described previously occur at slightly larger intermolecular
distances, with the angle $\theta_{\rm e}$ of the $C_{\rm s}$ minimum being
$\sim20^{\circ}$ smaller. These surfaces are all qualitatively similar to
Figure \ref{fig:int}a, except in the region $\theta=140^{\circ}$ to
$180^{\circ}$. A third minimum appears at $\theta=180^{\circ}$ for CH$_{3}$I +
Li: it is located at a similar intermolecular distance and is of comparable
depth ($\sim250$ cm$^{-1}$) to the $C_{\rm s}$ minimum on the same surface. All
geometries included in Figure \ref{fig:int} gave T1
diagnostics\cite{LTijqc1989} not exceeding $0.033$, indicating that
single-reference RCCSD(T) calculations are expected to be
reliable.\cite{RASjpca2000}

The potential energy surfaces involving the remaining alkali-metal atoms have
similar qualitative features for each methyl halide. We therefore focus on
quantitative comparisons between key stationary points on the sixteen surfaces.
The anisotropy of each surface can be inferred from the interaction energies
and geometrical parameters characterizing the stationary points. We consider
three stationary points on each surface, one at $\theta=0^{\circ}$, one at
$\theta=180^{\circ}$, and the third at the position of the $C_{\rm s}$ minimum.
By choosing to characterize only a few points on each surface, we are able to
perform calculations at a higher level of theory and allow for relaxation of
secondary geometrical parameters.

Stationary-point searches using RCCSD(T) are time-consuming and it is more
efficient to use RMP2 instead, since it produces similar geometrical parameters
for the systems of interest. As a benchmark example we examined the $C_{\rm s}$
minimum on the fixed-monomer CH$_{3}$F + Li surface, where the optimized
RCCSD(T)/AVDZ intermolecular parameters are $R=2.44$ \AA\ and
$\theta=147.0^{\circ}$ and the well depth is 1750 cm$^{-1}$ (see Figure
\ref{fig:int}a). Fixed-monomer optimizations performed using RMP2/AVDZ produced
similar values, with $R$ and $\theta$ larger by only 0.02 \AA\ and
0.6$^{\circ}$, respectively.

The fixed-monomer approximation is a good one for systems with weak
intermolecular forces, but there may be significant monomer distortions if the
interactions are comparable to monomer vibrational frequencies. Among the
systems of interest here, this effect is most significant for CH$_3$F + Li.
When the secondary geometrical parameters were also optimized during the
RMP2/AVDZ stationary point search for this system, the C-F bond stretched by
0.03 \AA\ and the associated $R$ and $\theta$ parameters differed from the
fixed-monomer RCCSD(T)/AVDZ results by 0.00 \AA\ and $-1.2^{\circ}$,
respectively. The RMP2/AVDZ//RCCSD(T)/AVDZ well is 514 cm$^{-1}$ deeper than in
the fixed-monomer calculation. We also obtained optimized $R$ and $\theta$
parameters using RMP2/AVTZ, which differed from the RCCSD(T)/AVDZ results by
$-0.03$ \AA\ and 0.1$^{\circ}$, respectively. From these tests optimizations at
the RMP2/AVDZ level were deemed adequate for our purposes. However, the AVTZ
basis set makes a significant difference to the final energetics, so we have
used it in the single-point RCCSD(T) calculations.

\begin{table*}[htp]
\caption{ \label{tab:opt} Characteristics of selected stationary points on the
ground-state nonreactive potential energy surfaces for the CH$_{3}X + A$
systems, computed at the RMP2/AVDZ//RCCSD(T)/AVTZ level of theory.
Counterpoise-corrected RMP2 energies were used to perform the stationary-point
searches. Binding energies ($D_{\rm e}$) are reported in cm$^{-1}$ with respect
to the energy of infinitely separated geometry-optimized monomers.
Intermolecular distances ($R_{\rm e}$) and angles ($\theta_{\rm e}$) are
reported in \AA\ and degrees, respectively. }
\begin{tabular*}{13cm}{
l c c c c c c c c c c c c c c c c c c c c c}
\hline
\hline
Alkali- &&\multicolumn{9}{c}{CH$_{3}{\rm F} \cdots  A$} && \multicolumn{9}{c}{CH$_{3}{\rm Cl} \cdots  A$}\\
\cline{3-11}\cline{13-21}
metal  &&\multicolumn{2}{c}{$\theta =$ 0$^{\circ}$} && \multicolumn{2}{c}{$\theta =$ 180$^{\circ}$} && \multicolumn{3}{c}{$C_{\rm s}$ minimum}
       &&\multicolumn{2}{c}{$\theta =$ 0$^{\circ}$} && \multicolumn{2}{c}{$\theta =$ 180$^{\circ}$} && \multicolumn{3}{c}{$C_{\rm s}$ minimum}\\
\cline{3-4}\cline{6-7}\cline{9-11}\cline{13-14}\cline{16-17}\cline{19-21}
atom   && $D_{\rm e}$ & $R_{\rm e}$
       && $D_{\rm e}$ & $R_{\rm e}$
       && $D_{\rm e}$ & $R_{\rm e}$ & $\theta_{\rm e}$
       && $D_{\rm e}$ & $R_{\rm e}$
       && $D_{\rm e}$ & $R_{\rm e}$
       && $D_{\rm e}$ & $R_{\rm e}$ & $\theta_{\rm e}$\\
\hline
Li && 78.9 & 5.90 && 1852  & 2.57 && 2073  & 2.44 & 145.8 && 109.7 & 6.16 && 166.1 & 4.85 && 846.0 & 2.79 & 115.3 \\
Na && 61.8 & 6.31 && 1815  & 3.02 && 1922  & 2.94 & 152.7 && 100.3 & 6.42 && 160.8 & 5.00 && 312.9 & 3.83 & 113.0 \\
K  && 57.4 & 6.61 && 511.2 & 3.60 && 644.9 & 3.42 & 145.3 &&  65.0 & 7.13 && 119.3 & 5.52 && 218.5 & 4.62 & 111.3 \\
Rb && 49.5 & 6.73 && 555.3 & 3.68 && 585.4 & 3.58 & 149.8 &&  55.6 & 7.12 && 106.2 & 5.77 && 224.2 & 4.62 & 111.9 \\
\hline
Alkali-&&\multicolumn{9}{c}{CH$_{3}$Br $\cdots$ $A$} && \multicolumn{9}{c}{CH$_{3}$I $\cdots$ $A$}\\
\cline{3-11}\cline{13-21}
metal &&\multicolumn{2}{c}{$\theta =$ 0$^{\circ}$} && \multicolumn{2}{c}{$\theta =$ 180$^{\circ}$} && \multicolumn{3}{c}{$C_{\rm s}$ minimum}
      &&\multicolumn{2}{c}{$\theta =$ 0$^{\circ}$} && \multicolumn{2}{c}{$\theta =$ 180$^{\circ}$} && \multicolumn{3}{c}{$C_{\rm s}$ minimum} \\
\cline{3-4}\cline{6-7}\cline{9-11}\cline{13-14}\cline{16-17}\cline{19-21}
atom   && $D_{\rm e}$ & $R_{\rm e}$
       && $D_{\rm e}$ & $R_{\rm e}$
       && $D_{\rm e}$ & $R_{\rm e}$  & $\theta_{\rm e}$
       && $D_{\rm e}$ & $R_{\rm e}$
       && $D_{\rm e}$ & $R_{\rm e}$
       && $D_{\rm e}$ & $R_{\rm e}$  & $\theta_{\rm e}$\\
\hline
Li && 135.0 & 6.31 && 259.0 & 4.46 && 785.2 & 2.89 & 105.0 && 166.0 & 6.42 && 469.4 & 4.22 && 663.8 & 3.23 & 94.5 \\
Na && 127.5 & 6.51 && 241.8 & 4.67 && 344.1 & 3.92 & 102.5 && 147.7 & 6.60 && 416.7 & 4.47 && 336.8 & 4.31 & 89.9 \\
K  &&  77.3 & 7.29 && 190.0 & 5.21 && 360.1 & 4.14 & 107.7 &&  56.6 & 7.85 && 317.3 & 5.02 && 376.5 & 4.31 & 97.1 \\
Rb &&  53.7 & 7.51 && 152.3 & 5.60 && 223.8 & 4.82 &  99.3 &&  31.7 & 7.98 && 260.1 & 5.28 && 182.3 & 5.23 & 88.6 \\
\hline
\hline
\end{tabular*}
\end{table*}

Interaction energies and geometrical parameters resulting from optimizations at
the RMP2/AVDZ//RCCSD(T)/AVTZ level of theory are reported in Table
\ref{tab:opt} for all sixteen CH$_{3}X + A$ systems. From these results a few
trends emerge. For the CH$_{3}$F systems, the global minima have $C_{\rm s}$
geometries and the well depth rises gently from $\theta_{\rm e}$ to
$180^\circ$. CH$_{3}$F + Li and CH$_{3}$F + Na have substantially deeper wells
than CH$_{3}$F + K and CH$_{3}$F + Rb. Most of the remaining complexes have
global minima at $C_{\rm s}$ geometries, but the preference for this geometry
over $\theta=180^\circ$ decreases from Cl to I, and CH$_{3}$I + Na and
CH$_{3}$I + Rb actually have global minima at $\theta=180^\circ$. For each of
CH$_{3}$Cl, CH$_{3}$Br and CH$_{3}$I, the interactions with Na, K and Rb are
comparable but that with Li is substantially stronger.

It is also important to consider the anisotropy around the molecular $C_{\rm
3v}$ axis. To investigate this effect in the region of the $C_{\rm s}$ minimum
for CH$_{3}$F + Li we have performed an RMP2/AVDZ//RCCSD(T)/AVTZ calculation
with the H-C-F-Li dihedral angle constrained to $\phi=0$. The values $R_{\rm
e}=2.44$ and $\theta_{\rm e}=145.8^{\circ}$ were obtained, which are identical
to those for $\phi=60^\circ$ (see Table \ref{tab:opt}) and the anisotropy is
$\sim 1$ cm$^{-1}$. In the region with $\theta<145.8^{\circ}$, the spatial
distribution of the hydrogen atoms causes more significant anisotropy about the
molecular $C_{\rm 3v}$ axis. RMP2/AVDZ//RCCSD(T)/AVTZ calculations were
performed to locate the $C_{\rm s}$ saddle point with $\phi=60^\circ$ and
$\phi=0$. We find that $R_{\rm e}$ and $\theta_{\rm e}$ shift from 5.90~\AA\
and $62.6^\circ$ at $\phi=60^\circ$ to $5.67$ \AA\ and 91.7$^{\circ}$ at
$\phi=0$, indicating that the position of the repulsive wall in this region
shifts significantly upon rotation about the molecular $C_{\rm 3v}$ axis.

It is useful to compare our results with those for other systems. The value of
2073 cm$^{-1}$ obtained here for the depth of the $C_{\rm s}$ entrance-channel
well in CH$_3$F + Li is similar to the value of 2100 cm$^{-1}$ obtained for the
HF + Li interaction.\cite{FLFjpca2013} Ref.\ \onlinecite{SZHfd2008} reported
well depths for the interactions between various alkali-metal atoms and the NH
molecule. For the lowest quartet state at linear $A$-NH geometries, they are
1799.1, 651.3, 784.7, and 709.3 cm$^{-1}$ for Li, Na, K, and Rb, respectively.
For $A$ + NH$_3$ systems,\cite{ZHpra2008} the corresponding well depths for the
ground state are 5104, 2359, 2161 and 1862 cm$^{-1}$, respectively. These
results follow the trend noted above for the interactions with CH$_{3}$Cl,
CH$_{3}$Br and CH$_{3}$I, where the well is similar magnitude for Na, K, and Rb
but substantially deeper for Li. The CH$_{3}$F systems are rather different,
since the wells of CH$_{3}$F + K and CH$_{3}$F + Rb are similar but those for
both CH$_{3}$F + Li and CH$_{3}$F + Na are considerably deeper.

\begin{figure}
\centering
\includegraphics[width=3.37in]{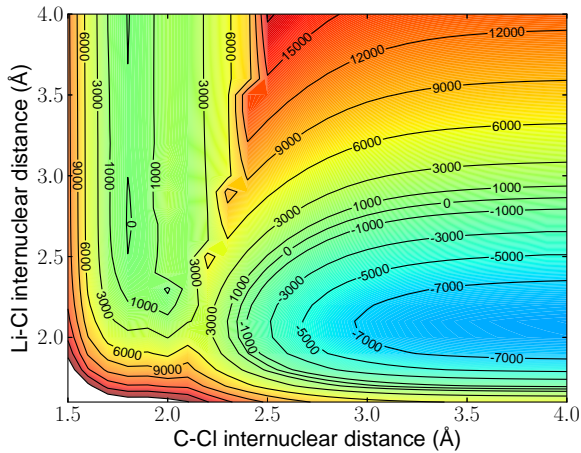} \\
\caption{CASSCF(11,10)/AVDZ potential energy surface for the CH$_{3}$Cl + Li
reaction as a function of the Li-Cl and C-Cl internuclear coordinates, with
with the Li-Cl-C angle fixed at 180$^\circ$ and other coordinates
optimized. Contours represent interaction energies in cm$^{-1}$.
\label{fig:cut1} }
\end{figure}
\begin{figure}
\centering
\includegraphics[width=3.37in]{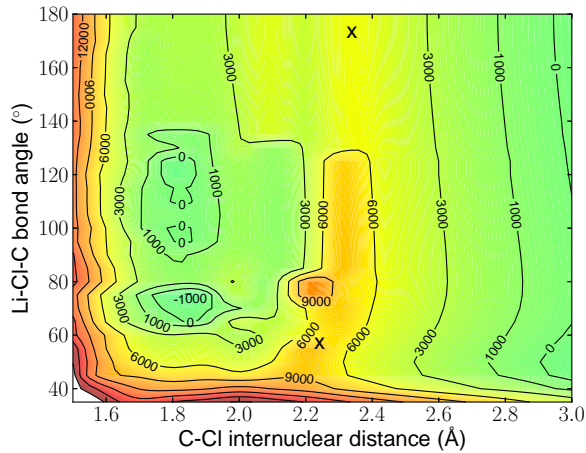}
\caption{CASSCF(11,10)/AVDZ potential energy surface for the CH$_{3}$Cl + Li
reaction as a function of the Li-Cl-C angle and C-Cl internuclear distance,
with the Li-Cl distance fixed at 2.75 \AA\ and other coordinates optimized.
Contours represent interaction energies in cm$^{-1}$.
\label{fig:cut2} }
\end{figure}

\subsection{Reactive potential energy surface for
${\rm CH}_3{\rm Cl+Li} \rightarrow {\rm CH}_3 + {\rm LiCl}$
}
\label{sec:3b}

The full potential energy surface for a CH$_{3}X + A$ reaction is a
hypersurface in 12 dimensions. However, in many cases a chemical reaction is
governed by only a few internal coordinates. In this section we examine
low-dimensional cuts through the reactive surface for the model system
CH$_{3}$Cl + Li. Interaction energies were computed at the CASSCF(11,10)/AVDZ
level of theory. Figure \ref{fig:cut1} shows the energy as a function of the
Li-Cl and C-Cl internuclear distances, with with the Li-Cl-C angle fixed at
180$^\circ$ and other coordinates optimized. Figure \ref{fig:cut1} shows the
entrance and exit channels, as well as the region of the transition state. The
entrance channel is centered about the equilibrium CH$_{3}$Cl bond distance
($R_{{\rm CCl}} \approx 1.8$~\AA\ for $R_{{\rm LiCl}}
> 2.3$~\AA) and the exit channel is centered about the equilibrium LiCl bond
distance ($R_{{\rm LiCl}} \approx 2.1$~\AA\ for $R_{{\rm CCl}} > 3.0$~\AA ).
The reaction is exothermic, with a late barrier. The transition state is
product-like and occurs at $R_{{\rm LiCl}} \approx 2.2$~\AA\ and $R_{\rm CCl}
\approx 2.2$~\AA.

In the vicinity of the transition state, there is an avoided crossing between
two electronic states of the same symmetry with quite different charge
distributions. This causes the ground-state adiabatic wave function to change
rapidly when passing through this region and, as a result, geometry
optimizations converge toward dissimilar relaxed nuclear configurations on each
side of the barrier. There is thus an energy cusp evident where the two
geometry-optimized surfaces meet, which is a result of the
reduced-dimensionality subspace of optimization parameters we have chosen to
work within. The height of the cusp is a lower bound to the true barrier
height. This is because it corresponds physically to the crossing of two
segments of the same adiabatic hypersurface which are connected in
higher-dimensional space.

Figure \ref{fig:cut2} shows the interaction energy as a function of the Li-Cl-C
bond angle and the C-Cl internuclear distance with the Li-Cl bond distance
fixed at 2.75 \AA. Here it may be seen that there are actually two low-energy
saddle points connecting reactants to products (marked by ``X'' on Figure
\ref{fig:cut2}). The energetically favored angles for reaction span the region
$\geq 140^{\circ}$ near $R_{\rm CCl}=2.4$ \AA, with the lowest barrier near
$180^{\circ}$. This corresponds to the classic rebound reaction. The second
pathway near $\theta_{\rm LiClC}=60^{\circ}$ and $R_{\rm CCl}=2.3$ \AA\ is
steep and narrow, and in fact it appears only if the geometry is optimized at
each point during the construction of the potential. This pathway corresponds
to insertion of the Li atom into the C-Cl bond. Since its barrier is higher
than that for the rebound reaction, we do not consider this pathway further.

\begin{figure}
\includegraphics[width=3.37in]{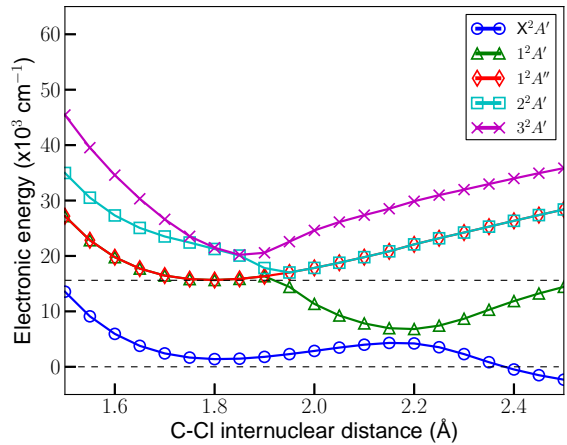}
\caption{
\label{fig:cpt}
The five lowest-lying CASSCF(3,6)//MS-MR-CASPT2(3,6)/6-31G$^{\ast}$ potential energy curves
along the C-Cl bond-breaking coordinate for CH$_{3}$Cl + Li $\rightarrow$ CH$_{3}$ + LiCl with
$R_{\rm LiCl}$ and $\theta_{\rm LiCCl}$ fixed at 2.25 \AA\ and 180$^{\circ}$, respectively.
The two dashed horizontal lines represent the separated reactant energies for Li($^2S$) and
Li($^2P$).
}
\end{figure}

If the potential energy surface in Figure \ref{fig:cut2} is traced along the
fixed angle $\theta=180^{\circ}$ from $R_{\rm CCl}=1.8$ to 3.0~\AA, there is a
double barrier between reactants and products. In order to investigate whether
this phenomenon persists when the potential is computed using higher levels of
theory, we have performed CASSCF(3,6)//MS-MR-CASPT2(3,6)/6-31G$^{\ast}$
calculations with $R_{{\rm LiCl}}$ fixed at 2.25 \AA, $\theta_{{\rm LiClC}}$
fixed at 180$^{\circ}$, and $R_{{\rm CH}}$ and $\theta_{{\rm ClCH}}$ optimized
for each point. Five contracted reference states were treated together to
obtain a balanced description of the avoided crossings. The resulting potential
curves are shown in Figure \ref{fig:cpt}, and shows a series of avoided
crossings as a charge-transfer state descends through a series of covalent
states with increasing $R_{{\rm LiCl}}$. The lowest potential curve shows only
one barrier, arising from an avoided crossing with this state. This provides
evidence that there is in reality only one barrier to reaction along the cut
with $\theta_{{\rm LiClC}}=180^{\circ}$. It also shows that there is at least
one excited state of the collision complex that lies below the energy of
CH$_{3}$Cl + Li($^2$P). Reactions involving excited alkali-metal atoms are in
general unlikely to have significant barriers.

\subsection{Reaction profiles and energetics for
${\rm CH}_3X+A \rightarrow {\rm CH}_3 + AX$
}
\label{sec:3c}
In the previous section we considered CH$_3$Cl + Li as a model to explore the
mechanism of the more general CH$_{3}X + A \rightarrow {\rm CH}_{3} + AX$ DCT
reaction. We found that the barrier to reaction is lowest when the alkali-metal
atom approaches head-on towards the halogen end of the methyl halide molecule,
i.e., with $\theta_{\rm CClLi}=180^{\circ}$. In this section we obtain
estimates of the activation energies of all sixteen DCT reactions.

In order to characterize a transition state fully, a saddle point must be
located on the PES, characterized by one imaginary vibrational frequency.
Despite a lengthy effort to locate saddle points for these DCT systems, the
searches never converged, and we were forced to develop a less rigorous
procedure for estimating the relevant activation energies. We will refer to the
quantities generated in this way as ``barrier heights'' in order to distinguish
them from true activation energies.

In our procedure we start from an optimized reactant van der Waals complex and
perform constrained optimizations along R$_{{\rm C}X}$, incrementally
stretching the C-$X$ bond and reoptimizing the secondary geometrical parameters
of the complex. The procedure is then repeated, this time starting from the
product van der Waals complex and incrementally compressing the C-$X$ bond
length while reoptimizing the other geometrical parameters. During these
calculations we fix $\theta_{{\rm LiClC}}=180^{\circ}$, constraining the
complex to $C_{\rm 3v}$ symmetry. This stepwise procedure is halted in each
direction when a calculation fails to converge. Reaction profiles were produced using
this procedure at the RMP2//RCCSD(T)/AVDZ level of theory since the MS-MR-CASPT2 method
is too expensive for routine calculations when Dunning's basis sets are employed.

\begin{figure}[t]
\includegraphics[width=3.37in]{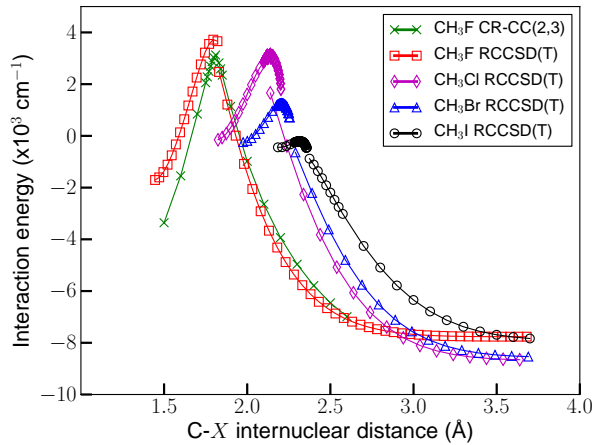}
\caption{
Reaction profiles for methyl halides CH$_{3}X$ with Li computed at the
RMP2//RCCSD(T)/AVDZ or CR-CC(2,3)/VDZ level of theory. Note that the forward
and reverse segments do not match perfectly in the transition state region due
to rapid variation of the geometry and charge density as described in the text.
\label{fig:bar}}
\end{figure}

The resulting potential energy curves are shown in Figure \ref{fig:bar}. We
include curves for all four of the systems involving Li to illustrate features
common to all 16 reactions studied. The forward and reverse potential energy
curve segments can be seen to match fairly well in the region of the transition
state. However, for every reaction we examined, only the forward segment
yielded a peak. For each system we determined the geometry at a point within 1
$\mu$Hartree of the peak of the forward potential curve. After this geometry
was obtained at the RMP2//RCCSD(T)/AVDZ level of theory, subsequent
single-point energy calculations were carried out to obtain barrier heights at
the RCCSD(T)/AVTZ level of theory.

\begin{table}[t]
\caption{Barrier heights for CH$_{3}X + A \rightarrow \rm{CH}_{3} + AX$
reactions. Structures were located at the RMP2//RCCSD(T)/AVDZ level of theory
using a numerical search method, as described in the text. Results of
single-point energy calculations performed on the resulting structures are
reported below, as computed at the RCCSD(T)/AV$x$Z level of theory.
\label{tab:bar} }
\begin{tabular}{l c c c c c
}
\hline
\hline
Alkali-metal &Basis set&   \multicolumn{4}{c}{Methyl halide molecule} \\
\cline{3-6}
Atom         &  level  & CH$_{3}$F & CH$_{3}$Cl & CH$_{3}$Br & CH$_{3}$I \\
\hline
Li     & AVDZ & 3728                & 3026                & 1163                & -249 \\
       & AVTZ & 4258                & 3398                & 1436                & -105 \\
       & AVQZ & 4227                & 3502                & 1467                & -124 \\
\hline
Na     & AVDZ & 4790                & 4063                & 1961                &  331 \\
       & AVTZ & 4615                & 4586                & 2290                &  567 \\
       & AVQZ & n/c\footnotemark[1] & 4743                & 2309                &  512 \\
\hline
K      & AVDZ & 6237                & 3706                & 2385                &  1133\\
       & AVTZ & 6328                & 4542                & n/c\footnotemark[1] &  1151\\
       & AVQZ & 6233                & 5544                & n/c\footnotemark[1] &  836 \\
\hline
Rb     & AVDZ & $>$6000             & 3810                & 2584                &  736 \\
       & AVTZ & n/c\footnotemark[1] & n/c\footnotemark[1] & 2971                &  1069\\
       & AVQZ & n/c\footnotemark[1] & n/c\footnotemark[1] & 2864                &  965 \\
\hline
\hline
\multicolumn{2}{c}{Semi-empirical values\footnotemark[2]}& 15000 & 4400 & 2000 & 200 \\
\hline
\hline
\end{tabular}
\footnotetext[1]{Calculations did not converge.}
\footnotetext[2]{Taken from Ref.\ \onlinecite{Wjpc1979}}.
\end{table}

To confirm that RCCSD(T) gives acceptable energetics in the region of the
transition state for these DCT reactions, we have also computed the CH$_{3}$F +
Li $\rightarrow$ CH$_{3}$ + LiF reaction profile at the CR-CC(2,3)/VDZ level of theory.
Augmented basis functions were not used in these calculations to reduce
the computational expense associated with numerical gradients. The
CR-CC(2,3)/VDZ reaction profile is also included in Figure \ref{fig:bar}. Its
peak is centered at $R_{\rm CF} \sim 1.81$ \AA\ with a reaction barrier height
of 3100 cm$^{-1}$. For comparison, the RMP2//RCCSD(T)/VDZ calculations gives a
peak near $R_{\rm CF} \sim 1.79$ with a reaction barrier height of 3500
cm$^{-1}$. The two methods give good agreement for the position of the barrier
and acceptable agreement ($\sim 500$ cm$^{-1}$) for the barrier height. Keeping
the magnitude of this discrepancy in mind, we proceed using
RMP2/AVDZ//RCCSD(T)/AV$x$Z to compute the remaining barrier heights for this
class of reactions.

Table \ref{tab:bar} reports barrier heights determined using this procedure for
all 16 systems and a variety of basis sets. The CH$_{3}X + A$ barrier heights
mostly increase with increasing alkali-metal atomic number and decrease with
increasing halogen atomic number. Some forward calculations involving K and Rb
did not reach a peak before failing to converge, so for these cases we provide
a lower bound for the height of the barrier. The AVQZ results for Li and Na
systems indicate that the AVTZ results are converged to within $\sim200$\
cm$^{-1}$ with respect to the basis-set size. The signs of barrier heights
suggest that activation barriers exist for all systems considered here except
CH$_{3}$I + Li, which has a submerged barrier.

The factor that limits the accuracy of the barrier heights presented in Table
\ref{tab:bar} is the reliability of the RCCSD(T) method in the region of the
transition state. Some of the T1 diagnostics near the peak geometries are as
large as $\sim$0.10, indicating that there is significant multi-reference
character in these regions. However, since the resulting potential energy
curves follow physical trends, we believe that the RCCSD(T) results still give
a good estimate of the barrier heights. A preliminary benchmark study comparing
RCCSD(T) with multireference configuration interaction calculations for the
related DEA reaction of CH$_{3}$F suggests that the error in the RCCSD(T)
activation barriers might be as large as $\sim 1000$ cm$^{-1}$. We therefore
cannot be certain that barriers exist for any of the four CH$_{3}$I + $A$
reactions.

Finally, it is interesting to compare our barrier heights with the values
obtained by Wu from the analogous dissociative electron attachment processes.
These are included in the bottom row of Table \ref{tab:bar}. Wu's estimates are
in reasonably good agreement with ours for systems involving CH$_{3}$Cl,
CH$_{3}$Br, and CH$_{3}$I. The DEA approach slightly underestimates activation
barriers for systems involving Li and overestimates barriers for systems
involving Na to Rb. For systems involving CH$_{3}$F, Wu's estimated reaction
barriers are 2 to 3 times larger than the values we obtain. This large
discrepancy is probably attributable to the significant stabilization of the
product-like transition state by the presence of the alkali-metal atom.

\section{Conclusions and Future Outlook}
\label{sec:4}
We have investigated both nonreactive and reactive potential energy surfaces
for interaction of methyl halides with alkali-metal atoms. Reactive collisions
occuring at cold and ultracold temperatures can usually proceed only if there
is an exothermic reaction pathway with a submerged or nonexistent barrier. Of
the 16 reactant combinations considered in this study, submerged barriers are
likely to be present only for the CH$_{3}$I + Li reactions, though they cannot
be ruled out for CH$_{3}$I with heavier alkali-metal atoms. For the remaining
12 atom-molecule combinations, significant barriers to reaction are predicted.

For the nonreactive interactions between methyl halides and alkali-metal atoms,
we find deep minima and strong anisotropies in the well region for all systems
considered. Systems involving Li have especially strong and anisotropic
interactions, but collision systems involving Li will also have larger
centrifugal barriers than for other alkali-metal atoms and these may suppress
cold inelastic collisions. In future work we will investigate the nonreactive
surfaces in greater detail and explore the extent to which centrifugal barriers
suppress inelastic collisions of trapped methyl halide molecules with ultracold
Li atoms.

\section*{Acknowledgments}

We are grateful to Heather Lewandowski for discussions that identified this
problem and to the Engineering and Physical Sciences Research Council for
funding under grant no.\ EP/I012044/1.

\end{document}